\documentclass[aps,pre,twocolumn,groupedaddress]{revtex4}
\usepackage{graphicx}

\begin{document}

\title{Oscillations in DC driven "barrier" discharges:\\
numerical solutions, stability analysis and phase diagram}

\author{Danijela D. \v{S}ija\v{c}i\'c$^1$, Ute Ebert$^{1,2}$ 
and Ismail Rafatov$^{1,3,4}$}

\affiliation{$^1$CWI, P.O.Box 94079, 1090 GB Amsterdam, The Netherlands,}
\affiliation{$^2$Dept. Physics, Eindhoven Univ. Techn., The Netherlands,}
\affiliation{$^3$American University -- Central Asia, Bishkek, Kyrgyzstan}
\affiliation{$^4$Middle East Technical University, Ankara, Turkey}

\date{\today}

\newcommand \be{\begin{equation}}
\newcommand \ee{\end{equation}}
\newcommand \ba{\begin{eqnarray}}
\newcommand \ea{\end{eqnarray}}
\def\nn{\nonumber}
\def\np{\newpage}
    
\begin{abstract} 
A short gas discharge layer sandwiched with a semiconductor layer 
between planar electrodes shows a variety of spatio-temporal patterns. 
The paper focusses on the spatially homogeneous spontaneous oscillations
while a DC voltage is applied; the results on these homogeneous oscillations
apply equally to a planar discharge in series with any resistor with 
capacitance. We define the minimal model, identify its independent 
dimensionless parameters and then present results of the full time-dependent
numerical solutions of the model as well as of a linear stability analysis
of the stationary state. Full numerical solutions and the results
of the stability analysis agree very well. The stability analysis
is then used for calculating bifurcation diagrams. 
We find semi-quantitative agreement with experiment for the
diagram of bifurcations from stationary to oscillating solutions as well as 
for amplitude and frequency of the developing limit cycle oscillations.
\end{abstract}
\maketitle

\section{Introduction}

Gas discharges on the transition from Townsend to glow regime
exhibit a wealth of spatio-temporal structures. Besides striations,
i.e., longitudinal waves in a long 
discharge column \cite{Jonas2,Bruhn,Golub,Letellier,Bultel}, short discharges
with wide lateral aspect ratio can also exhibit rich spatio-temporal
structures in the transversal direction as reported by a number of
authors \cite{Gwinn,Islamov,Dong,Nasuno}.
This is even the case when the externally applied voltage 
is stationary and the gas is pure, as long as the system is sandwiched
between planar electrodes and at least one Ohmic layer. 
An interesting sequence of experiments has been performed 
in M\"unster \cite{Str,privatStr} where the bifurcations 
between different spatio-temporal states in parameter space
were investigated very systematically.

As in our previous paper \cite{PRL}, we focus in the present one
on the purely temporal oscillations that occur in a spatially
homogeneous mode. This focus has two reasons: first, understanding
the temporal structures is a first systematic step towards 
understanding the full spatio-temporal structures; second, 
there are numerous observations of temporal oscillations 
in comparable parameter regimes 
\cite{Zoran,Phelps93II,Zoran3,PhelpsPRE97,Fiala,Pitch,Astrov,Kolo04}.
For the oscillations, the setup need not contain an Ohmic layer
as in \cite{Str,privatStr}, a resistor with capacitance in the circuit
will have the same effect on the gas discharge.

In the previous paper \cite{PRL}, we concentrated on
the question whether a simple two-component reaction-diffusion
model for current and voltage in the gas discharge layer would
be sufficient to describe the oscillations. Such a model
is suggested through similarities with patterns formed in
a number of physical, chemical or biological systems like the 
Belousov-Zhabotinski reaction, Rayleigh-Benard convection, patterns
in bacterial colonies, in Dictyostelium or in nerval tissue etc.
However, the actual results of a realistic gas discharge model 
are in conflict with a simple two-component reaction diffusion 
approximation that neglects the height and subsequent memory 
of the system. This can be seen, in particular, from the occurence
of a period doubling cascade as well as from
analytical model reductions \cite{PRL}. Similar period doubling cascades are observed experimentaly in \cite{pd1,pd2,pd3,pd4,pd5,pd6}.

In the present paper, we continue the analysis of the full
gas discharge model, coupled to a high-Ohmic layer and driven by
a stationary voltage. The focus is now on quantitative comparison 
with experiment, on a stability analysis and on the derivation of bifurcation diagram.
 The specific experiment
to be analyzed was performed in nitrogen at 40 mbar within a gap
of 0.5 or 1 mm wide while the semiconductor was a layer of 1.5 mm
photosensitively doped GaAs. To the whole structure, voltages in the range
of 500 to 800 V were applied.
As in our previous papers \cite{us,us2,PRL},
we restrict the analysis to the direction normal to the layers,
hence assuming homogeneity in the transversal directions.
The experimental system actually shows a transition from
a homogeneous stationary to a homogeneous oscillating state,
and the theory presented here reproduces essential features of these 
experiments. At the same time, the investigation serves as a 
gauge point for a later analysis of spatio-temporal patterns.

In detail, we define the model as a set of partial differential 
equations and perform a dimensional analysis 
in Section II. In Section III, first the physical parameters
and the numerical details of solving the PDE's in time are given.
Then qualitative and quantitative results of numerical solutions
and experiments are discussed. In particular, the hysteresis
between stationary and oscillating solutions is demonstrated
numerically, amplitude and frequency of the limit cycle oscillations
as a function of applied voltage and conductivity of the semiconductor
are compared with experimental results, and the physical mechanism of
the oscillation is discussed. In Section IV, it is explained
how the stability analysis about a stationary solution of the complete
system is performed. In Section V, the results of the stability
analysis are presented. First a convincing agreement between
numerical solutions of the full PDE's and the stability analysis 
results is found. Then the stability analysis is used to calculate
bifurcation diagrams for the transition from stationary to oscillating
states that are then compared with experiment.
The paper concludes with Section VI.

%\newpage

\section{The model}

The experiment consists of two layers, a gas discharge and a semiconductor,
sandwiched between two planar electrodes to which a DC voltage is applied.
In this section, the equations are defined and a dimensional analysis
is performed to identify the independent parameter combinations of the 
problem. This also serves to identify physical processes and time scales.

\subsection{Gas discharge layer}

In the gas discharge, two ionization mechanisms cooperate to
maintain conductivity: the so-called $\alpha$ process of impact 
ionization in the bulk of the discharge, and the $\gamma$ process 
of secondary emission at the cathode. The classical ``fluid'' 
approximation consists of continuity equations for electron density
$n_e$ and positive ion density $n_+$, coupled to the Poisson equation 
for the electric field $E$:
\ba
\label{1}
\partial_t\;n_e \;+\; \partial_r J_e
&=& source ~,
\\
\label{2}
\partial_t\;n_+ \;+\; \partial_r J_+
&=& source ~,
\\
\label{3}
\partial_r E &=& {{\rm e}\over{\varepsilon_0}} \;(n_+ -n_e)~. 
\ea
The spatial coordinate $r$ is normal to the layers, and in the present paper,
it is assumed that there are no variations in the transversal directions.
The gas is assumed to be non-attaching, i.e., no negative ions are formed.
Also photo-ionization, Ohmic heating, nonlocal interactions and diffusion
are neglected in this simplest approximation.
The particle current densities ${J}_e$ and $ J_+$ are approximated
by a drift motion that is linear in the field
\ba
\label{4}
{J}_e = - n_e \;\mu_e \;{ E}~,~~~
{J}_+ =  n_+ \;\mu_+ \;{ E} ~. 
\ea
The source term on the right hand side of Eqs.\ (\ref{1}) and (\ref{2})
is approximated by impact ionization in the classical Townsend form
\ba
\label{6}
source = |n_e \mu_e { E}| \;\alpha_0
\;\mbox{\large{e}}^{\textstyle -E_0/|{ E}|} ~.
\ea

The one-dimensional approximation of Eqs.\ (\ref{1}), (\ref{2}) and (\ref{3})
makes the total electric current $J(t)$ homogeneous
\be
\label{13}
\epsilon_0\partial_t  E(r,t)+{\rm e}J_e(r,t)+{\rm e}J_+(r,t)=J(t)~~~,~~~
\partial_rJ(t)=0 .
\ee
This identity can be used to substitute $J_e$ or $J_+$ by $J(t)$.
In the present analysis, we will keep $n_e(r,t)$ and $E(r,t)$ 
as independent fields and express $n_+$ and $J_+$ by these fields 
and the total current $J(t)$. 

The model is completed by boundary conditions on the electrode.
At the anode which is located at $r=0$, electrons are absorbed and ions
are absent:
\be
\label{G5}
J_+(0,t)=0~~~\Longleftrightarrow~~~n_+(0,t)=0~.
\ee
At the cathode which is located at $r=d$, impacting ions can 
liberate electrons by secondary emission with rate $\gamma$:
\be
\label{G6}
|J_e(d,t)|=\gamma \;|J_+(d,t)|~~~\Longleftrightarrow~~~
\mu_e n_e(d,t)=\gamma\mu_+ n_+(d,t)~.
\ee
Note that consistenly with \cite{us,PRL}, but in contrast with
most other literature, the anode is on the left hand side at $r=0$.
This has the advantage that the electric field is positive,
and sign mistakes when evaluating $E$ or $|E|$ cannot occur.

Substantial densities of charged particles change the electric field
according to (\ref{3}), and the electric field determines drift 
and ionization rates of the particles according to Eqs. (\ref{1}),
(\ref{2}), (\ref{4}) and (\ref{6}). Therefore the process is nonlinear 
as soon as space charges become relevant. It causes the well-known 
transition from the linear Townsend discharge to the nonlinear glow discharge.

\subsection{Semiconductor layer and complete circuit}

The semiconductor layer of thickness $d_s$ is assumed to have
a homogeneous and field independent
conductivity $\sigma_s$ and dielectricity constant $\epsilon_s$:
\be
\label{S1}
J_s(t)=\sigma_s { E}_s(t)~,~~~
\label{MX1}
q=\epsilon_s\epsilon_0 \; \partial_r {  E}~.
\ee
As there are no space charges in the bulk of the semiconductor,
the electric field is homogeneous, and
voltage and field are related through
$U_s(t)={ E}_s(t)d_s$. The equation of charge conservation 
$\partial_t q+\partial_r J_s=0$ in one dimension leads again 
to the homogeneity of the total current density $J(t)$
\be
\label{S2}
\epsilon_s\epsilon_0\partial_t { E}_s(t) + J_s(t) = J(t) ,
\ee
that is the same as in the gas discharge (\ref{13}).
Hence in macroscopic parameters, the semiconductor solves
\ba
\label{S3}
C_s\partial_tU_s(t)+J_s(t)=J(t)~~&,&~~U_s(t)=R_sJ_s(t) ,
\\
C_s=\frac{\epsilon_s\epsilon_0}{d_s}\quad~~~&,&~~~\quad
R_s=\frac{d_s}{\sigma_s} .
\ea
where $C_s$ is the capacitance per area.

According to (\ref{S3}), perturbations of $U_s(t)$ or $J_s(t)$
decay on the Maxwell time scale 
\be
\label{Ts}
T_s=C_s R_s=\frac{\epsilon_s\epsilon_0}{\sigma_s}.
\ee
This time scale is independent of the thickness of the
semiconductor layer although it represents the time that the charge 
needs to cross it. The time scale of the experimentally
observed oscillations is of the order of $T_s$, and therefore also 
approximately proportional to $1/\sigma_s$ as will be discussed 
in Section III.D. 

Actually, for the present investigation of one-dimensional oscillations,
the specific structure of a planar semiconductor layer is not required,
but any serial component of the electric circuit with capacitance $C_s$
and resistance $R_s$ will support the same equation (\ref{S3}).

The total stationary voltage $U_t$ over the complete system is 
\be
\label{Ug}
U_t=U(t)+U_s(t)~,~~U(t)=\int_0^{d}E(r,t)dr~~,~~\partial_tU_t=0.
\ee
According to (\ref{S3}) and (\ref{Ug}), the dynamics of the voltage $U(t)$
on the gas discharge obeys the equation
\be
\label{Ug2}
T_s\partial_t U=U_t-U(t)-R_s J(t).
\ee

\subsection{Dimensional analysis and system definition}

The dimensional analysis is performed as previously in 
\cite{PREuwc,us,us2,PRL}.
We introduce the dimensionless coordinates and fields
\ba
\label{DimA}
&& z=\frac{r}{X_0}~~,~~\tau=\frac{t}{t_0}~~,~~
 \sigma(z,\tau)=\frac{n_e(r,t)}{n_0} , 
\\
&&{\cal E}(z,\tau)=\frac{E(r,t)}{E_0}~~,~~~
{\cal U}=\frac{U}{E_0X_0} ~~,~~
j=\frac{J}{{\rm e}n_0X_0/t_0},
\nn
\ea
measuring quantities in terms of the intrinsic parameters
of the system
\be
X_0=\frac1{\alpha_0}~~,~~
t_0=\frac1{\alpha_0\mu_eE_0}
~~,~~~~n_0=\frac{\epsilon_0\alpha_0E_0}{\rm e}.
\ee
After eliminating the ion dynamics by the total current $j(\tau)$,
the equation of motion of the gas discharge becomes
\ba
\label{g1}
\partial_\tau\sigma&=&\partial_z j_e
+j_e \alpha({\cal E})~~,~~j_e=\sigma{\cal E} ,\\
\label{g2}
\partial_\tau {\cal E}&=&j(\tau) -(1+\mu) j_e - 
\mu {\cal E} \partial_z {\cal E} ,
\ea
and the boundary conditions (\ref{G5}) and (\ref{G6}) read
\ba
\label{g3}
\partial_\tau {\cal E}(0,\tau)&=&j(\tau) - j_e(0,\tau ) ,\\
\label{g4}
\partial_\tau {\cal E}(L,\tau)&=&j(\tau) - 
\frac{1+\gamma}{\gamma} j_e(L,\tau ) .
\ea
The intrinsic dimensionless parameters of the gas discharge are
the mobility ratio $\mu$ of electrons and ions and the length ratio $L$
of system size and inverse cross section of impact ionization
\be
\mu=\frac{\mu_+}{\mu_e}~~~,~~~
L=\frac{d}{X_0} .
\ee
The discharge is coupled to the semiconductor and the DC voltage source
${\cal U}_t$ through (\ref{S3}) as
\ba
\label{s2}
\tau_s\partial_\tau{\cal U}(\tau)
&=&{\cal U}_t-{\cal U}(\tau)-{\cal R}_s j(\tau) ,
%\\
%\label{s3I}
%{\cal U}(\tau)&=&\int_0^L{\cal E}(z,\tau)\;dz
%j(\tau)=\kappa_s\;
%\Big(\;{\cal U}_t-{\cal U}(\tau)-\tau_s\partial_\tau{\cal U}(\tau)\;\Big) ,
\ea
with the dimensionless parameters
\be
\tau_s=\frac{T_s}{t_0} .~~~,~~~
{\cal R}_s=\frac{R_s}{E_0t_0/({\rm e}n_0)} .
\ee
The voltage ${\cal U}(\tau)=\int_0^L{\cal E}(z,\tau)\;dz$ 
is related to the electric field ${\cal E}$ and potential $\phi$
in differential form as
\be
\label{s3}
{\cal E}(z,\tau)=-\partial_z \phi(z,\tau)~~,~~ 
{\cal U}(\tau)=\phi(0,\tau)-\phi(L,\tau) ,
\ee
where gauge freedom allows one to choose 
\be
\label{s4}
\phi(0,\tau)=0.
\ee

Hence the dynamics of the complete system is described by Eqs.\
(\ref{g1})--(\ref{g4}), (\ref{s2}), (\ref{s3}) and (\ref{s4}). 
The system is characterized completely
by the independent dimensionless parameters $\mu$, $L$ and $\gamma$
for the gas discharge layer, $\tau_s$ and ${\cal R}_s$ for 
the semiconductor layer and the total applied DC voltage ${\cal U}_t$.

%\newpage

\section{Numerical solutions of the dynamics}

In this section, this dynamical model is solved numerically
and the results are compared with experiments. 
We discuss physical parameters under A and numerical details under B.
In C, qualitative features of experimental and numerical system
are compared like the bistability between stationary and oscillating
state. In D, a quantitative comparison between theory and experiment 
is performed, and the dependence of amplitude and frequency
of the oscillation as a function of ${\cal U}_t$ and $1/{\cal R}_s$
is determined numerically. Finally, in E, we discuss the mechanism 
of the oscillations and identify the surface charge effects 
that are inherent in our model.

\subsection{Physical parameters}

In the experiment \cite{Str}, nitrogen at a pressure of 40 mbar
was used in gaps with widths of 0.5 or 1 mm.
The article \cite{Str} contains mainly data for the 0.5 mm
gap, while the Ph.D. thesis \cite{privatStr} also contains 
more data for 1 mm. The gas discharge was coupled to 
a semiconductor layer of GaAs with a width of $d_s=$~1.5 mm
and a dielectricity constant $\epsilon_s=13.1$.
Through photosensitive doping, the conductivity of the 
semiconductor layer could be increased by about an order of magnitude;
the dark conductivity was $\sigma_s=3.2\cdot 10^{-8} 
(\Omega {\rm cm})^{-1}$. For the discharge gap of 0.5 mm width, 
voltages in the range of 500 to 600~V were used; 
for the gap of 1 mm width, the applied voltages were 
in the range of 580 to 740~V.

%The experimentally measured Paschen curves in \cite{privatStr}
%that show the breakdown voltage $U$ of the gas discharge as a function 
%of pressure times gap width $pd$, can be used to determine 
%parameters of the discharge experimentally like in particular
%the parameters $\alpha_0$, $E_0$ and $\gamma$ of the ionization reaction.
%The classical model predicts that the Paschen curves for different system
%sizes should be indistinguishable. In practice, they do not precisely
%fall on top of each other. Nevertheless, we use them for approximating
%the values of $\gamma$ and $E_0$.

Of course, the predictive power of the theory depends on the model
approximations as well as on the chosen parameters. Our simple 
classical model will not give fully quantitative agreement. 
On the other hand, its simple structure and few parameters 
give a chance of physical understanding and control.

For the gas discharge, we used the ion mobility 
$\mu_+=23.33 \;\mbox{cm}^2\mbox{/Vs}$ and electron
mobility $\mu_e=6666.6 \;\mbox{cm}^2\mbox{/Vs}$. 
For $\alpha_0=Ap=[27.78 \mu{\rm m}]^{-1}$
and $E_0=Bp=10.26 \;\mbox{kV/cm}$,
the value from \cite{Raizer} was used. The gap widths of
$d=$~0.5 and 1~mm then correspond to dimensionless gap widths
$L=$~18 and 36. 
For $\gamma$, we used the value 0.08 determined from experimental
Paschen curves in \cite{privatStr}. It should be noted that
our classical model predicts that the Paschen curves 
(i.e., the breakdown voltage $U$ of the gas discharge as a function 
of pressure times gap width $pd$) for different system
sizes should be indistinguishable. In practice, they do not precisely
fall on top of each other. 

\begin{figure}[htbp]
  \begin{center}
    \includegraphics[width=0.49\textwidth]{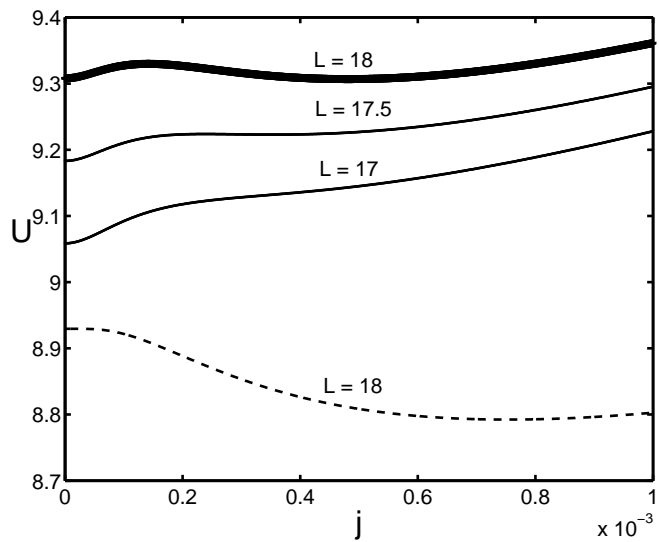}\\
    \caption{Current-voltage characteristics for $\gamma=0.08$
(solid lines) and $\gamma=0.1$ (dashed line) for the dimensionless
gap widths $L$ as indicated in the figure.}
  \end{center}
\end{figure}

It is interesting to note how sensitive the theoretical results
are to small changes of the secondary emission coefficient $\gamma$,
in particular, for the short gap with $L=18$. This is illustrated in Fig.~1. 
The upper three solid lines show the 
shape of the current voltage characteristics for $\gamma=0.08$ and
gap widths of $L=$ 17, 17.5 and 18. As discussed in more detail in
\cite{us,us2}, the characteristics can be supercritical ($L=$~17, positive
differential conductivity for all values of the current $j$), mixed II
($L=$~17.5, Townsend breakdown voltage lower than the local voltage minimum
for $j\ne0$) or mixed I ($L=$~18, Townsend breakdown voltage higher 
than the local voltage minimum for $j\ne0$). The dashed line shows the 
characteristics for $L=$18 and $\gamma=0.1$. ${\cal U}$ then overall
is considerably lower and the characteristics is fully subcritical, 
i.e., the voltage has only one minimum as function of current $j$ 
and this occurs for a value $j\ne0$. This subcritical behavior 
corresponds to the classical textbook case where the characteristics 
bends down from the Townsend breakdown voltage towards a voltage
minimum in the glow discharge regime --- as we have discussed in
\cite{us,us2} in detail, this requires a sufficiently large
system size. For $\gamma=0.08$, the characteristics becomes subcritical
for system size $L>L_{crit}=e^2\ln \big[ (1+\gamma)/\gamma\big]=19.2$ while
the transition to supercritical behavior is determined numerically
\cite{us} to the value of $L=17.2$.

Data on the coefficient $\gamma$ of secondary electron emission
are relatively scarce, so it is quite common \cite{applPhys} to use 
it as an adjustable parameter as we do. The tabulated data for $\alpha_0=Ap$ 
and $E_0=Bp$ from \cite{Raizer} together with the Paschen curve 
for $d=$~0.5 mm from \cite{privatStr} would suggest 
$\gamma=0.03$, but that would mean that the characteristics would be
supercritical up to $L=24.9$, then it would develop some regime with 
negative differential conductivity, and it would become subcritical 
only for $L>L_{crit}(\gamma=0.03)=26.1$.

We conclude that the gap with width 0.5 mm (corresponding to $L=18$)
is so sensitive to the not very well known parameter $\gamma$
that an analysis of the experimental data would be rather uncertain.
Furthermore, the approximation of purely local interactions becomes
worse in shorter gaps. Finally, the electric fields in short discharges
are higher and vary more; therefore the assumption that $\gamma$
does not depend on $E$ becomes more restrictive.
For this reason, we chose to analyze the system with gap width 1 mm
$(L=36)$.

We recall that the following intrinsic scales 
\ba
X_0\approx 27.78\;\mu\mbox{m} ~~&,&~~ 
t_0\approx 40.6\cdot10^{-12} \;\mbox{s},
\nn\\
n_0 \approx 2.04 \cdot10^{12} / \mbox{cm}^3~~&,&~~ 
E_0\approx 10.26 \;\mbox{kV/cm}
\ea
enter the dimensional analysis (\ref{DimA}).
Therefore the dimensionless parameters for a system with
gap width of $d=$~1~mm and applied voltages in the range 
from 500 to 740 V are in our simulations:
\ba
\label{param}
&&\mu=0.0035~~,~~L=36~~,~~\gamma=0.08,
\nn\\
&& \tau_s=0.243~{\cal R}_s~~,~~3 \cdot 10^5\le{\cal R}_s\le3 \cdot 10^6
\nn\\
&&17.5\le{\cal U}_t\le26.
\ea
Here, the dimensionless capacitance of the semiconductor layer is 
${\cal C}_s = 0.243 $, and its dimensionless 
characteristic time scale is $\tau_s={\cal C}_s{\cal R}_s$.
The value ${\cal R}_s=3 \cdot 10^6$ for the semiconductor resistance
corresponds to the
dark conductivity of $\sigma_s=3.2\cdot10^{-8}/(\Omega{\rm cm})$, 
and ${\cal R}_s=3 \cdot 10^5$ corresponds to the fully photo-activated 
conductivity $\sigma_s=3.2\cdot10^{-7}/(\Omega{\rm cm})$.
The dimensionless voltage range of $17.5\le{\cal U}_t\le26$
corresponds to the dimensional range of 500 V~$\le U_t\le$~740 V.

\subsection{Numerical solution strategy}

Equations (\ref{g1})--(\ref{s4}) were solved numerically
with an implicit temporal discretization, which makes the calculation
numerically stable for arbitrary time and space steps.
After discretization, the dynamical equations (\ref{g1}) and (\ref{g2}) 
have the form
\ba
\frac{\sigma_{i}^{m+1}-\sigma_{i}^{m}}{\Delta\tau}&=&
\frac{(\sigma\mathcal{E})_{i+1}^{m+1}-(\sigma\mathcal{E})_{i}^{m+1}}{\Delta
z}+\big(\mathcal{E}\sigma\;\alpha({\cal E})\big)_{i}^{m+1},
\nonumber\\
\frac{\mathcal{E}_{i}^{m+1}-\mathcal{E}_{i}^{m}}{\Delta\tau}
&=&j^{m}-\mu\mathcal{E}_{i}^{m}\frac{\mathcal{E}_{i}^{m+1}-
\mathcal{E}_{i-1}^{m+1}}{\Delta z}
-(1+\mu)\left(\mathcal{E}\sigma\right)_{i}^{m},
\nonumber\\
\ea
where $i$ parametrizes the spatial and $m$ the temporal grid.

For known $\sigma^{m}$ and $\mathcal{E}^{m}$ at time step $m$,
the boundary condition on the left (\ref{g3}) determines 
\be
\mathcal{E}_{1}^{m+1}= \mathcal{E}_{1}^{m}+\Delta
\tau\big(j^{m}-\left(\mathcal{E}\sigma\right)^{m}_{1}\big),
\ee 
then the other fields $\mathcal{E}^{m+1}_{i}$ are calculated
successively from the left to right ($i=2,3,..,N$) by the equation
\begin{equation}
\mathcal{E}_{i}^{m+1} =
\frac{\mathcal{E}_{i}^{m}\left(1+\frac{\mu\Delta \tau}{\Delta
z}\mathcal{E}_{i-1}^{m+1} -(1+\mu)\Delta
\tau\sigma_{i}^{m}\right)+\Delta \tau
j^{m}}{1+\frac{\mu\Delta \tau}{\Delta
z}\mathcal{E}_{i}^{m}}.
\end{equation}
For $\sigma_i^{m+1}$, the boundary condition on the right (\ref{g4})
determines 
\begin{equation}
\sigma_{N}^{m+1}=
\left(j^{m}-\frac{\mathcal{E}_{N}^{m+1}-\mathcal{E}_{N}^{m}}{\Delta
\tau}\right)/\left(\frac{1+\gamma}{\gamma}\mathcal{E}_{N}^{m+1}\right).
\end{equation}
The remaining $\sigma_{i}^{m+1}$ can now be calculated successively 
from the right to left ($i=N-1,N-2,..,1$) as
\be
\sigma_{i}^{m+1}= \frac{\sigma_{i}^{m}+\frac{\Delta \tau}{\Delta
z} (\sigma\mathcal{E})_{i+1}^{m+1}}{1+\frac{\Delta
\tau}{\Delta z} \mathcal{E}_{i}^{m+1} -\Delta \tau
\mathcal{E}_{i}^{m+1}\alpha(\mathcal{E}_{i}^{m+1})}.
\ee
The total current $j^m$ in these equations is determined by
\ba
j^{m}=\frac1{{\cal R}_s+\tau_{s}L}
\Bigg[\mathcal{U}_t-\mathcal{U}^{m}
+\tau_s\Bigg(\frac{\mu}{2}\left(\left(\mathcal{E}^{m}_{N}\right)^{2}
-\left(\mathcal{E}^{m}_{1}\right)^{2}\right)&&
\nonumber\\
+(1+\mu)\Delta
z\sum_{i=1}^{N-1}(\mathcal{E}\sigma)^{m}_{i}\Bigg)\Bigg].&&
\ea
This identity can be derived from (\ref{s2}) where $\partial_\tau{\cal U}$
is identified with $\int_0^L dz\;\partial_\tau{\cal E}$ through
(\ref{s3}), and then for $\partial_\tau{\cal E}$, the identity
(\ref{g2}) is used.

The results presented in Figures 2 to 9 are derived on a grid with
$\Delta z=36/600$ and $\Delta \tau=180/600$ which gives a sufficient 
numerical accuracy. 

\subsection{Qualitative features of experimental and numerical oscillations:
hysteresis amd limit cycles}

The experiments \cite{Str} show approximately periodic oscillations.
They are quite anharmonic with long phases of low current interrupted
by a short current pulse. Depending on applied voltage ${\cal U}_t$ 
and resistance of the semiconductor layer ${\cal R}_s$, either
the homogeneous stationary or the homogeneous oscillating state 
are dynamically stable. Inbetween, there is a regime of bistability
where it depends hysteretically on the previous state whether the
system is stationary or oscillating.

The same qualitative behavior can be observed in our numerical solutions.
First, the upper panel in Fig.~2 shows the current $j(\tau)$ as a function 
of time for the system with the parameters from (\ref{param}) and 
${\cal R}_s=4 \cdot 10^5$ and ${\cal U}_t=19.5$ (which corresponds
to $\sigma_s=2.4\cdot 10^{-7}/(\Omega{\rm cm})$ and $U_t=555$~V). 
After some transient, the current relaxes to periodic unharmonic 
oscillations. The lower panel in Fig.~2 shows the voltage ${\cal U}(\tau)$ 
over the gas discharge; the voltage on the semiconductor is correspondingly 
${\cal U}_t-{\cal U}(\tau)$. In dimensional units, the peak current 
of the oscillations is about 9~mA/cm$^2$ and the frequency is about
120~kHz.

\begin{figure}[htbp]
  \begin{center}
    \includegraphics[width=0.49\textwidth]{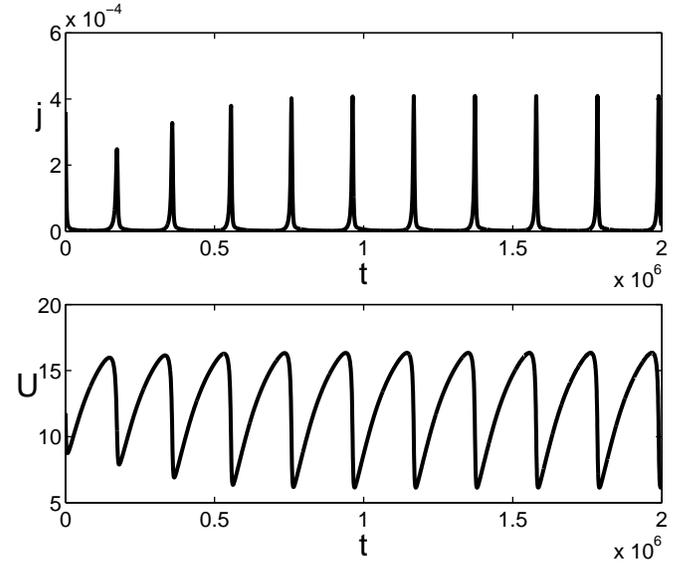}\\
    \caption{$j(\tau)$ and ${\cal U}(\tau)$ for the parameters from 
        (\ref{param}), ${\cal R}_s=4 \cdot 10^5$ and ${\cal U}_t=19.5$.}
  \end{center}
\end{figure}

The same numerical data for current $j$ and voltage ${\cal U}$ are
shown as a phase space plot in Fig.~3.
The figure shows more precisely the approach to a limit cycle.
Fig.~3 contains two additional lines, namely the current voltage
characteristics of the gas discharge ${\cal U}={\cal U}(j)$
and the load line ${\cal U}={\cal U}_t-{\cal R}_sj$. Their intersection
marks the stationary solution of the system. In the present case,
it is located in the low current regime close to the Townsend limit,
while the peak current explores the regime of subnormal glow.

\begin{figure}[htbp]
  \begin{center}
    \includegraphics[width=0.49\textwidth]{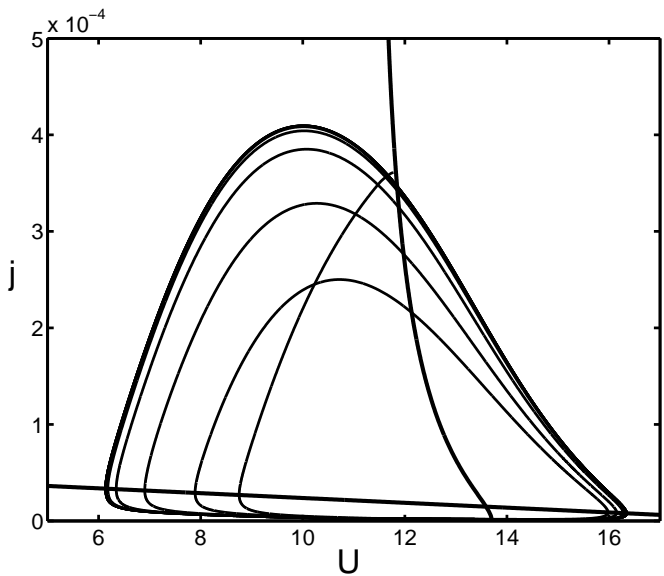}\\
    \caption{Phase space plot of the data from Fig.~2. 
      After some transient time, a stable limit cycle is reached.
      Also drawn are the current-voltage-characteristics ${\cal U}={\cal U}(j)$
      of the gas discharge and the load line ${\cal U}={\cal U}_t-{\cal R}_sj$.
      Their intersection denotes the stationary solution.}
  \end{center}
\end{figure}

The system of Figs.~2 and 3 is actually in the bistable regime.
For different initial conditions that are a sufficiently small perturbation
of the stationary state, the same system relaxes to the stationary point.
This is shown as phase space plot in Fig.~4.

\begin{figure}[htbp]
  \begin{center}
    \includegraphics[width=0.49\textwidth]{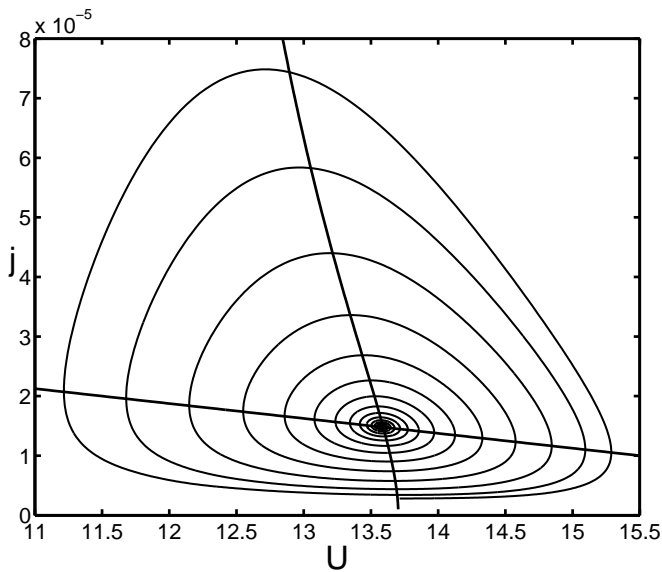}\\
    \caption{System with exactly the same parameters as in Figs.~2 and 3,
    but for different initial conditions. The system now spirals inwards
    towards the stationary point.}
  \end{center}
\end{figure}

If the applied voltage ${\cal U}_t$ becomes large enough,
the stationary state becomes unstable for any initial condition.
The search for appropriate parameters was guided by the stability
analysis described in sections IV and V of this paper. We find
that ${\cal U}_t=24$ ($U_t=684$~V) with all other parameters unchanged can be
used as an example of a system where the stationary solution is dynamically
unstable, and the system runs away from this initial state
and eventually reaches a limit cycle oscillation. This behavior is 
shown in Fig.~5 as $j(\tau)$ and ${\cal U}(\tau)$,
while Fig.~6 shows the corresponding phase space plot.

\begin{figure}[htbp]
  \begin{center}
    \includegraphics[width=0.49\textwidth]{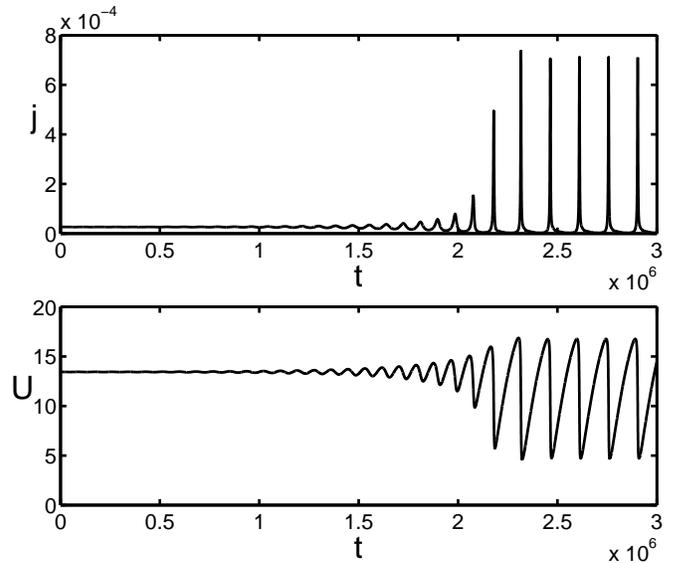}\\
    \caption{$j(\tau)$ and ${\cal U}(\tau)$ for the parameters from 
        (\ref{param}), ${\cal R}_s=4 \cdot 10^5$ and ${\cal U}_t=24$.
        The stationary state now is linearly unstable and develops
        into a limit cycle.}
  \end{center}
\end{figure}

\begin{figure}[htbp]
  \begin{center}
    \includegraphics[width=0.49\textwidth]{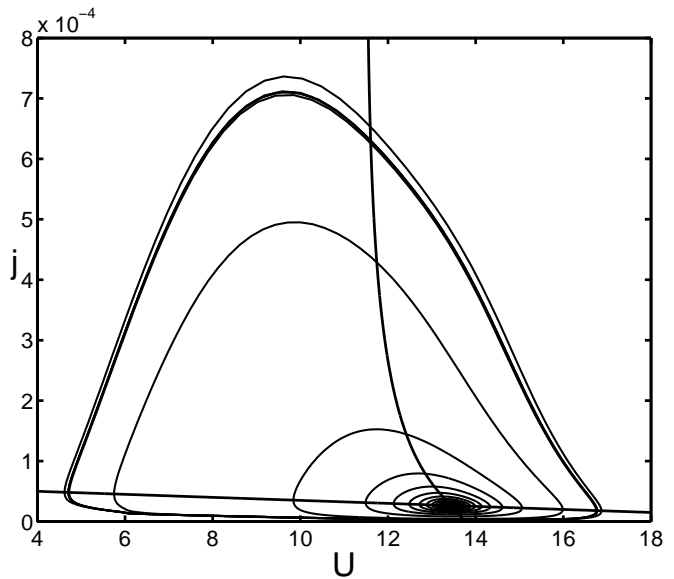}\\
    \caption{Phase space plot of the data from Fig.~5 with 
      current-voltage-characteristics and load line.}
  \end{center}
\end{figure}

%\newpage

%$\;$

%\newpage

\subsection{Quantitative comparison: amplitude and frequency of oscillations}

The qualitative agreement of numerical solutions and experiment
now encourages a more quantitative comparison. The thesis \cite{privatStr}
contains diagrams on how frequency and maximal current amplitude 
depend on the semiconductor conductivity for a gas gap of 1 mm.
It also contains the remark that frequency and amplitude for fixed
conductivity depend in about the same way on the applied voltage 
as in the 0.5 mm gap of Ref.~\cite{Str}.

The same diagrams can also be derived from the numerically obtained
limit cycle oscillations, they are presented in Fig.~7. The figure shows
the current amplitude $A$ and frequency $f$ as a function of
semiconductor conductance $1/{\cal R}_s$ for fixed voltage 
${\cal U}_t$ or as a function of ${\cal U}_t$ for fixed $1/{\cal R}_s$.

\begin{figure}[htbp]
  \begin{center}
    \includegraphics[width=0.49\textwidth]{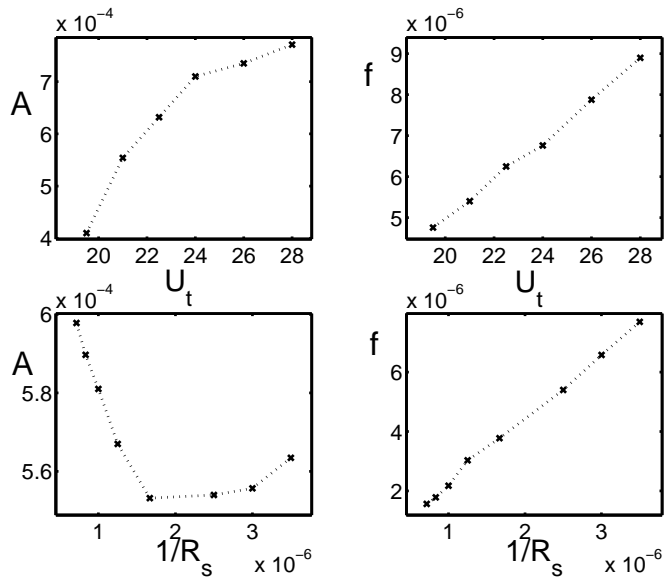}\\
    \caption{Amplitude $A$ and frequency $f$ of the current oscillations 
     as a function of applied total voltage ${\cal U}_t$ (for fixed 
     resistance $R_s=4 \cdot 10^5$) and as a function of conductivity 
     $1/{\cal R}_s$ (for fixed voltage ${\cal U}_t= 21$).}
  \end{center}
\end{figure}

We now compare the results. %of simulation and experiment.
The upper left panel shows that the maximal current amplitude $A$
as a function of applied voltage ${\cal U}_t$ is increasing
with decreasing slope. This agrees with the statements
written in \cite{privatStr}. The upper right panel shows
that the frequency $f$ is an almost linearly increasing function
of applied voltage ${\cal U}_t$, this is actually in contradiction
with the statement in \cite{privatStr} that the function would decrease.

The lower two panels allow a more quantitative comparison
since corresponding experimental diagrams can be found in
\cite{privatStr}. The experiments explore the range of
$0.6\cdot10^{-7}/(\Omega{\rm cm})\le\sigma_s\le 
2.8\cdot10^{-7}/(\Omega{\rm cm})$ which corresponds to 
$0.62\cdot10^{-6}\le 1/{\cal R}_s\le 2.9\cdot10^{-6}$.
The experimental diagrams for $U_t=605$~V 
and 616 V in \cite{privatStr} show, that the amplitude $A$ is very 
sensitive to this change while the frequency $f$ is rather robust. 
The numerical results are derived for ${\cal U}_t=21$ which
corresponds to $U_t=600$~V. 

In detail, the experimental curve for the current amplitude
for 605 V shows first an increase from 0.2 to 0.8 mA
with a subsequent sudden drop to essentially 0 from which
the current suddenly jumps to values from 1.0 to 1.5 mA.
For 616 V, in contrast, an almost continuous increase 
from 0.2 to 2.7 mA is observed for the same resistance range.
Not too suprisinngly, our numerical results reproduce neither
of these widely differing results at quite similar voltage. 
Rather, we observe an almost constant
value in the range of $5.5\cdot10^{-4}$ to $6.0\cdot10^{-4}$
in the lower left pannel.

On the other hand, for the variation of the frequency $f$ with
conductivity, experiments \cite{privatStr} both for
605 V and for 616 V observe an about linear
increase from 115 kHz or 125 kHz to 220 kHz 
($4.6\cdot10^{-6}\le f\le 8.8\cdot10^{-6}$ in our dimensionless units)
in the range of $0.62\cdot10^{-6}\le 1/{\cal R}_s\le 2.9\cdot10^{-6}$.
Our numerical results in this range of $1/{\cal R}_s$ show the same
linear increase, from $1.5\cdot10^{-6}$ to $6.5\cdot10^{-6}$.
We believe that this agreement is quite convincing, in particular,
since no parameter fitting was tried.

Summarizing, we find convincing agreement with experiment
for $A$ as function of ${\cal U}_t$ as well as for $f$ as a function of
$1/{\cal R}_s$. For the last, the available experimental results 
allow to identify an almost quantitative agreement.
The sensitivity of the experimental results
on $A$ as a function of $1/{\cal R}_s$ does not allow quantitative
comparison, and our results for $f$ as a function of ${\cal U}_t$
deviate in their functional form from the available statements 
about experimental results.

%\newpage

\subsection{Mechanism of the oscillations, reaction-diffusion models
   and surface charge}

The voltage profiles ${\cal U}(\tau)$ in Figs.~2 and 5 show 
that there are two processes involved in the oscillations. 

The first process occurs on the slow time scale $\tau_s$ of the 
semiconductor. It describes the exponential decay of the voltage 
${\cal U}_t-{\cal U}(\tau)-{\cal R}_sj \propto e^{-\tau/\tau_s}$
over the semiconductor layer according to Eq.~(\ref{s2}), as
long as the contribution of ${\cal R}_sj$ does not vary substantially. 
The decay time $\tau_s$ is 
the Maxwell time due to resistance and capacitance of the semiconductor layer.
$\tau_s$ accounts for the slow rise of the voltage ${\cal U}(\tau)$ 
over the gas discharge layer to a value above the current voltage 
characteristics of the gas discharge.

The other process is the electric breakdown of the gas discharge
layer for sufficiently large ${\cal U}(\tau)$ which leads to a 
current pulse and a rapid subsequent decay of ${\cal U}(\tau)$.
 
It has been suggested by a number of authors 
\cite{KGM,Rade89,Rade90,Rade92,Zoran3,Petro97,Fiala,Islamov,Muenster2}
that the current could be approximated by a similarly simple equation 
of the type $\partial_\tau j=g({\cal U},j)$, where $g$ vanishes on 
the current-voltage-characteristics. This would bring the
equations into a reaction diffusion form. However, as we already
have discussed in \cite{PRL}, such an approximation of the underlying
equations (\ref{g1})--(\ref{g4}), (\ref{s3}) and (\ref{s4}) is 
not possible, since it would not admit the period doubling
events observed in \cite{PRL}, and it would not allow the phase space
plots in Figs.~3, 4 and 6 to intersect the characteristics with a
nonvanishing derivative, as they definitely do.

The physical reason for this behavior is the finite response
time of the gas discharge layer, its ``inertia'' which doesn't allow
an instantaneous reaction of the current.
If ions are created by bulk impact ionization close to the
anode, they will cross the whole gap until they reach the cathode
and possibly liberate more electrons by secondary emission.
The time that the ions need to cross the gap, is therefore
an important scale of internal memory of the gas discharge.
It can be approximated as $\tau_{ion}\approx L/(\mu\:{\cal E})
\approx  L^2/(\mu\:{\cal U}(\tau))$
where $|{\cal E}|$ is some average field within the gas gap.
For the gap of $L=36$ ($d=1$~mm), the ion crossing time is estimated
as $2.6\cdot 10^4$ for ${\cal U}(\tau)=14$ or as $1.5\cdot 10^4$ 
for ${\cal U}(\tau)=24$ (which corresponds to 0.6 or 1 $\mu$s
in dimensional units). This time is of the
same order or larger than the duration of a current pulse, 
both in our numerical solutions and in the experimental results 
of Fig.~5 in \cite{Str}. (For the experiments on the 0.5 mm gap
of Fig.~4 in \cite{Str}, the situation seems to be different.)

Finally, it has been suggested in \cite{PurwinsNew} that the surface charge
on the interface between gas and semiconductor could play an important
role, in a similar way as in AC discharges. This is certainly true,
but the surface charge $q(\tau)$ is not an independent variable.
Rather it is fully determined by the solution discussed above through
\be
q(\tau)=\epsilon_s\;\frac{{\cal U}_t-{\cal U}(\tau)}{L}-{\cal E}(L,\tau).
\ee
The assumption that this surface charge is the only relevant charge
in the whole system doesn't lead to a satisfactory description either,
but the space charges in the gas discharge layer have to be taken into
account, too.

%\newpage

\section{Stability analysis: method}

The direct numerical solution of the dynamical problem is 
a time consuming procedure, that does not allow the exploration
of a wide set of parameter values. We therefore have developed 
a linear stability analysis of the stationary state. It determines 
whether the stationary state is dynamically unstable and how small 
perturbations of such a state grow. In the present section, 
we present the method, and in the following one the results.

\subsection{Problem setting and stationary solutions}

The dynamical equations from section II.C are summarized as 
\ba
\label{g10}
\partial_{\tau}\sigma&=&\partial_z j_e
+j_e \alpha({\cal E})~~,~~j_e=\sigma{\cal E} ,
\\
\label{g20}
\partial_{\tau} {\cal E}&=&j(\tau) -(1+\mu) j_e - 
\mu {\cal E} \partial_z {\cal E} ,
\\
\label{s20}
\tau_s\partial_{\tau}{\cal U}(\tau)
&=&{\cal U}_t-{\cal U}(\tau)-{\cal R}_s j(\tau) ,
\\
\label{s30}
0&=&\partial_z \phi(z,\tau)+{\cal E}(z,\tau)
\ea
with the boundary conditions
\ba
\label{g30}
\partial_{\tau} {\cal E}(0,\tau)&=&j(\tau) - j_e(0,\tau ) ,
\\
\label{g40}
\partial_{\tau} {\cal E}(L,\tau)&=&j(\tau) - 
\frac{1+\gamma}{\gamma} j_e(L,\tau) ,
\\
\phi(L,\tau)&=&-{\cal U}(\tau) ~~,~~
\label{s40}
\phi(0,\tau)=0.
\ea

The stationary solutions form the starting point of the perturbation
analysis. They solve the equations
\ba
\label{gs01}
\partial_zj_{e0}&=&-j_{e0}\;\alpha({\cal E}_0) ,\\
\label{gs02}
\mu {\cal E}_0\;\partial_z{\cal E}_0&=& j_0-(1+\mu)j_{e0} ,\\
\label{gs03}
\partial_z\phi_0&=&-{\cal E}_0 ,\\
\label{ss01}
{\cal U}_0&=&{\cal U}_t-{\cal R}_s j_0 
\ea
with boundary conditions
\ba
\label{bb01}
j_{e0}(0)&=&j_0~~~,~~~\frac{1+\gamma}{\gamma}\;j_{e0}(L) =j_0 ,
\\
\label{bb02}
\phi_0(0)&=&0~~~,~~~\phi_0(L)=-{\cal U}_0 .
\ea
Eqs.~(\ref{gs01})--(\ref{gs03}) with (\ref{bb01}) and (\ref{bb02})
define the current voltage characteristics
${\cal U}={\cal U}(j)$ of a stationary discharge in the regime
between Townsend and glow discharge \cite{Raizer,us,us2}. Eq.~(\ref{ss01})
is the load line due to the external circuit. The intersection of
load line and characteristics defines a generically discrete number
of stationary solutions of the system as a whole.

\subsection{Linear perturbations}

For linear perturbations about this stationary state,
we use the ansatz
\ba
\label{an1}
j_e(z,\tau)&= &j_{e0}(z) + j_{e1}(z)\; e^{\lambda\tau} , \\
{\cal E}(z,\tau)& =& {\cal E}_0(z)+ {\cal E}_1(z)\; e^{\lambda\tau} , \\
\phi(z,\tau)&=& \phi_0(z) +  \phi_1(z)\; e^{\lambda\tau} , \\
j(\tau) &= &j_0 + j_1\; e^{\lambda\tau} .
\ea
The lower index $0$ denotes the unperturbed stationary solutions while
the lower index $1$ denotes the linear perturbations about 
this stationary solution. The factorization of the perturbation
into a $z$ dependent function times the exponential $e^{s\tau}$
anticipates the eigenvalue problem of the solution.

In terms of the original variables, the explicit expansion
in first order perturbation theory is a lengthy expression, 
but in terms of the variables 
\be
h = \frac{\sigma_1{\cal E}_0+\sigma_0{\cal E}_1}{\sigma_0{\cal E}_0}
=\frac{j_{e1}(z)\;e^{s\tau}}{j_{e0}(z)}~~ 
\mbox{and} ~~~ g = {\cal E}_0 \;{\cal E}_1
\ee
the equations have a more compact form
\ba
\label{feq}
\partial_z h &=& \frac{\lambda}{{\cal E}_0} h - 
\left(\frac{\alpha'({\cal E}_0)}{{\cal E}_0} 
+ \frac{\lambda}{{\cal E}_0^3} \right)\;g ,\\
\label{feq2}
\partial_z g &=& -(1+\mu)\;\frac{j_{e0}}{\mu}\;h 
- \frac{\lambda}{\mu\;{\cal E}_0}\;g+\frac{j_1}{\mu},\\
\label{feq3}
\partial_z j_1 &=& 0 ,\\
\label{feq4}
\partial_z \phi_1 &=& -\frac{1}{{\cal E}_0}\; g 
\ea
with boundary conditions
\ba
\label{fbc1}
\phi_1 (0) &=& 0  , \\
\label{fbc2}
j_1 &=& \frac{\lambda}{{\cal E}_0(0)}\; g(0) + j_0\;h(0), \\
\label{fbc3}
j_1 &=& \frac{\lambda}{{\cal E}_0(L)}\; g(L) +  j_0\;h(L), \\
\label{fbc4}
{\cal R}_s j_1 &=& (1+\lambda\tau_s)~\phi_1 (L).
\ea
Here the equation $\partial_z j_1 = 0$ for the conservation
of the total current is written explicitly in order to bring 
the equations into the homogeneous form
\be
\label{matrix}
\partial_z
\left( \begin{array}{c}
h \\
g \\
j_1 \\
\phi_1 \\
\end{array} \right) = %M_s(z)
\left( \begin{array}{cccc}
 \frac{\lambda}{ {\cal E}_0} & - \Big(\frac{\alpha'}{{\cal E}_0} 
+ \frac{\lambda}{{\cal E}_0^3} \Big) & 0 & 0 \\ 
 -\frac{1+\mu}{\mu}\;j_{e0} & - \frac{\lambda}{\mu{\cal E}_0} 
& \frac{1}{\mu} & 0 \\
0 & 0 & 0 & 0 \\
0 & -\frac{1}{{\cal E}_0} & 0 & 0 \\
\end{array} \right)
\cdot
\left( \begin{array}{c}
h \\
g \\
j_1 \\
\phi_1 \\
\end{array} \right)
\ee

The boundary conditions (\ref{fbc1}) and (\ref{fbc2}) at $z=0$
can be written as orthogonality relations
\be
\label{bc0}
\left( \begin{array}{c} 
j_0 \\ \frac{\lambda}{{\cal E}_0(0)} \\ -1 \\ 0 \\
\end{array} \right)
\cdot
\left( \begin{array}{c} 
h \\ g \\ j_1 \\ \phi_1 \\
\end{array} \right)_0 
= 0
~~,~~
\left( \begin{array}{c} 
0 \\ 0 \\ 0 \\ 1 \\
\end{array} \right)
\cdot
\left( \begin{array}{c} 
h \\ g \\ j_1 \\ \phi_1 \\
\end{array} \right)_0 
= 0 .
\ee
The general solution $\vec{v}(z)$ of (\ref{matrix}) is therefore 
a superposition of two independent solutions $\vec{v}_1(z)$ 
and $\vec{v}_2(z)$ of (\ref{matrix}) that both obey (\ref{bc0}) in $z=0$:
\be
\label{sol}
\vec{v}(z) =
\left( \begin{array}{c}
h(z) \\ g(z) \\ j_1(z) \\ \phi_1(z) \\
\end{array} \right)
=C_1\;\vec{v}_1(z)+C_2\;\vec{v}_2(z) .
\ee
As initial conditions, one can choose, e.g.,
\be
\label{sol12}
\vec{v}_1(0)=
\left( \begin{array}{c}
1/j_0 \\ 0 \\ 1 \\ 0 \\
\end{array} \right) \; \;, \; \;
\vec{v}_2(0) = 
\left( \begin{array}{c}
0 \\ \frac{{\cal E}_0(0)}{\lambda} \\ 1 \\ 0 \\
\end{array} \right) .
\ee
The components of the two solutions are denoted as
$\vec{v}_i(z)=\big(h_i(z),g_i(z),j_{1,i}(z),\phi_{1,i}(z)\big)$.

The boundary conditions (\ref{fbc3}) and (\ref{fbc4}) at $z=L$
also have the form of orthogonality relations
\be
\label{bcL}
\left( \begin{array}{c} 
j_0 \\ \frac{\lambda}{{\cal E}_0(L)} \\ -1 \\ 0 \\
\end{array} \right)
\cdot
\left( \begin{array}{c} 
h \\ g \\ j_1 \\ \phi_1 \\
\end{array} \right)_L 
= 0
~~,~~
\left( \begin{array}{c} 
0 \\ 0 \\ - {\cal R}_s \\ 1+\lambda\tau_s \\
\end{array} \right)
\cdot
\left( \begin{array}{c} 
h \\ g \\ j_1 \\ \phi_1 \\
\end{array} \right)_L 
= 0 .
\ee
Now each one of these two conditions determines the ratio $C_1/C_2$ 
of the general solution (\ref{sol}):
\ba
&&C_1~\left[\textstyle
j_0 h_1(L)+\frac{\lambda}{{\cal E}_0(L)} g_1(L)-j_{1,1}(L)
\right] \nn \\ 
\label{cl1}
&&+ C_2~\left[\textstyle
j_0 h_2(L)+\frac{\lambda}{{\cal E}_0(L)} g_2(L)-j_{1,2}(L)
\right] = 0 , \\
&&C_1~\left[-{\cal R}_s j_{1,1}(L)+(1+\lambda\tau_s) \phi_{1,1}(L)\right] \nn \\
&&+C_2~\left[-{\cal R}_s j_{1,2}(L)+(1+\lambda\tau_s) \phi_{1,2}(L)\right] = 0 ,
\label{cl2}
\ea
where $j_{1,1}(L)=1=j_{1,2}(L)$, since these components have this
value at $z=0$ according to (\ref{sol12}):
$j_{1,1}(0)=1=j_{1,2}(0)$, and since the equation
of motion for $j_1$ is $\partial_zj_1=0$.
A nontrivial solution of both (\ref{cl1}) and (\ref{cl2}) 
requires the determinant
\ba
\label{det}
&&\Delta= \\
&&\left| \begin{array}{cc}
j_0 h_1(L)+\frac{\lambda}{{\cal E}_0(L)} g_1(L)-1
&
j_0 h_2(L)+\frac{\lambda}{{\cal E}_0(L)} g_2(L)-1
\\
-{\cal R}_s+(1+\lambda\tau_s)\phi_{1,1}(L)
&
-{\cal R}_s+(1+\lambda\tau_s)\phi_{1,2}(L)
\end{array} \right|
\nn
\ea
to vanish. This condition leads to a quadratic equation 
for the eigenvalue $\lambda$.

\subsection{Rescaling with $\mu$ and numerical calculation}

The eigenvalue $\lambda$ can now be calculated numerically. 

First, it should be noted, that the equation of motion
(\ref{matrix}) has matrix elements of very different size,
since $\mu$ is a very small parameter.
However, this apparent stiffness of the problem can be
removed by introducing the new parameters
\ba
\label{resc}
&&\iota_e=\frac{j_e}{\mu}~~,~~\iota=\frac{j}{\mu}~~,~~r_s={\cal R}_s\mu,\\
&&\bar\tau_s=\tau_s\mu~~,~~s=\frac{\lambda}{\mu}.
\nn
\ea
The introduction of rescaled current density and time scale
and resistivity has a direct physical motivation.
Previous analysis of the stationary solutions \cite{Raizer,us,us2} 
as well as the dynamical solutions of Section III and \cite{PRL}
show that velocities should
actually be measured on the time scale of the ions and not of the 
electrons. So the time scale should be measured in units of
$t_+=1/(\alpha_0\mu_+E_0)=t_0/\mu$ rather than in units of
$t_0=1/(\alpha_0\mu_eE_0)$. The rescaling (\ref{resc}) directly
follows from this consideration.

Now the eigenvalue $s$ can be calculated numerically as follows:
First an initial estimate $s_0$ is chosen. Then the two initial conditions
(\ref{sol12}) at $z=0$ are integrated numerically with (\ref{matrix}) 
up to $z=L$. Generically, the determinant $\Delta$ (\ref{det}) will then be 
non-vanishing. The request that the determinant does vanish, fixes
a new value for $s$ that is used for the next step of the iteration
within an under-relaxation method that garantuees the stability 
of the convergence. This procedure is repeated until an accuracy of
\be
\left| \frac{s_{k+1}-s_k}{s_{k+1}}\right| < 10^{-6}  
\ee
is reached.

The eigenvalue $s$ is in general a complex parameter whose real part 
describes the growth or decay of the oscillation amplitude
while its imaginary part describes the oscillation frequency.
Since $s$ is a parameter in the equation of motion (\ref{matrix}),
also the vector $\vec{v}(z)$ has complex entries. Therefore
16 real functions Re~$h_1(z)$, Im~$h_1(z)$ etc.\ have to be integrated
over $z$. It is convenient to also integrate the two real functions
$j_{e0}$ and ${\cal E}_0$ that enter the matrix (\ref{matrix}) together
with the perturbations.
The iteration program is written in fortran 90 with complex variables.
For the integration of equations, a 4th order Runge-Kutta method is used.
The number of grid points used was 500, since 1000 or 2000 grid points
give essentially the same result.

\section{Stability Analysis: Results}

In the present section, the validity of the stability analysis results 
are confirmed by comparison with numerical solutions of the full 
dynamical problem. The stability analysis is then used to determine 
the phase diagram for the onset of oscillating solutions. These phase
diagrams are then compared with experimental results, again with
semi-quantitative agreement.

\subsection{The structure of the results}

The stability analysis determines not only the complex eigenvalue $\lambda$,
but also the whole linear correction
\be
\label{65}
\vec{v}(z)=C_1 \left[\vec{v}_{1}(z) - 
\frac{{\cal R}_s+(1+\lambda\tau_s)~{\cal U}_{1,2}(L)}
{{\cal R}_s+(1+\lambda\tau_s)~{\cal U}_{1,1}(L)} \; \vec{v}_2(z)\right] ,
\ee
up to the arbitrary complex constant $C_1$.

This $\vec{v}(z)$ determines the evolution of current and voltage
in linear approximation about the stationary solution $(j_0,{\cal U}_0)$:
\ba
j(\tau) &=& j_0 + j_1\; e^{\lambda\tau}~+c.c. ,\\
{\cal U}(\tau) &=& {\cal U}_0 +  {\cal U}_1\; e^{\lambda\tau}~+c.c. ,
\ea
where c.c.\ denotes the complex conjugate. The ratio between ${\cal U}_1$ 
and $j_1$ is fixed through the boundary condition (\ref{fbc4}) 
to the value
\ba
\label{2bcU}
{\cal U}_1 &=& - \frac{j_1}{(1+\lambda\tau_s)/{\cal R}_s}
=r\;e^{i\alpha}\;j_1 ,
\\
\label{ralpha}
\mbox{where} &&  r = \frac{{\cal R}_s}{|1+\lambda\tau_s|}
~~\mbox{and}~~\\
&&
\cos\alpha=-\;\frac{1+{\rm Re}~\lambda\tau_s}{|1+\lambda\tau_s|}~~,
~~\sin\alpha=\frac{{\rm Im}~\lambda\tau_s}{|1+\lambda\tau_s|}.
\nn
\ea
The final result is
\ba
j(\tau) &=& j_0 + c\;\mu\;e^{{\rm Re}\;\lambda\tau}~
\cos({\rm Im}\;\lambda\tau+\alpha_0)~, \\
\label{pertU}
{\cal U}(\tau) &=& {\cal U}_0 -c~r\; e^{{\rm Re}\;\lambda\tau}~
\cos({\rm Im}\;\lambda\tau + \alpha+ \alpha_0)
\ea
where amplitude $c$ and absolute phase $\alpha_0$ reflect the arbitrary
factor $C_1$ in (\ref{65}) and are adjustable
while all other parameters are fixed.

\subsection{Comparison with solutions of the full PDE's}

As a check of accuracy, these solutions are now first compared
with numerical solutions of the full PDE problem.

For the set of parameters from Figs.~2, 3 and 4, the stationary 
solution is $(j_0,{\cal U}_0)=(1.49\cdot 10^{-5},13.583)$, 
and the eigenvalue $\lambda$ has the complex value 
$\lambda=-2.913 \cdot 10^{-6} \pm i~ 4.822 \cdot 10^{-5} $. 
As $\tau_s=340/\mu$ and ${\cal R}_s=1400/\mu$, 
the ratio of current and voltage amplitude $r=295/\mu$ and the phase 
shift $\alpha=98.69^o$ are determined through Eq.~(\ref{ralpha}).

The comparison of these predictions from the stability analysis
with numerical solutions of the full PDE problem are shown in
Fig.~8. Here the free parameters for the total amplitude $c$ and 
the absolute phase $\alpha_0$ were chosen such as to fit 
the PDE-data well.

\begin{figure}[htbp]
  \begin{center}
    \includegraphics[width=0.49\textwidth]{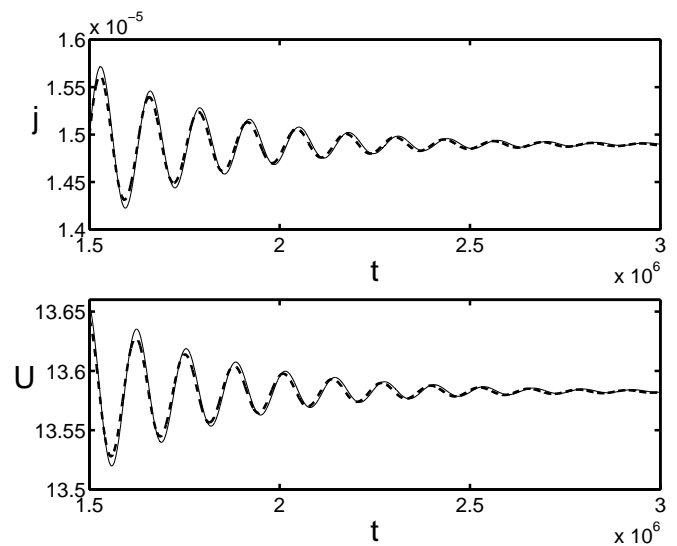}\\
    \caption{Comparison of $j(\tau)$ and ${\cal U}(\tau))$ 
from the stability analysis (solid lines) 
with the result from the simulation (dashed lines) 
for the parameter values of Figs.~2-4.}
  \end{center}
\end{figure}

This visual agreement can be tested in more detail. 
In particular, we used the PDE-data in the time interval 
$5 \cdot 10^5 < \tau < 6.5 \cdot 10^5$ to determine the phase 
shift $\alpha$ between ${\cal U}_1$ and $j_1$. It is 
$\alpha=(100 \pm 0.4)^o$, convincingly close to the predicted 
value of $\alpha=98.69^o$.

Increasing the total applied voltage ${\cal U}_t$, the real part
of the eigenvalue $\lambda$ grows until it becomes positive. This means
that the stationary solution becomes linearly unstable 
and perturbations will grow. An example of such behavior occurs
for ${\cal U}_t=24$ with all other parameters as before.
The stationary solution is then $(j_0,{\cal U}_0)=(2.64 \cdot 10^{-5}, 
13.441)$, the eigenvalue is $\lambda=2.493\cdot10^{-6}\pm 
i~7.375\cdot10^{-5}$, the ratio of current and voltage amplitude 
is $r=192/\mu$ and the phase shift is $\alpha=99.83^o$.

Fig.~9 shows again the comparison between these results and 
the numerical solutions of the full PDE's. Again, the agreement
is very convincing.

\begin{figure}[htbp]
  \begin{center}
    \includegraphics[width=0.49\textwidth]{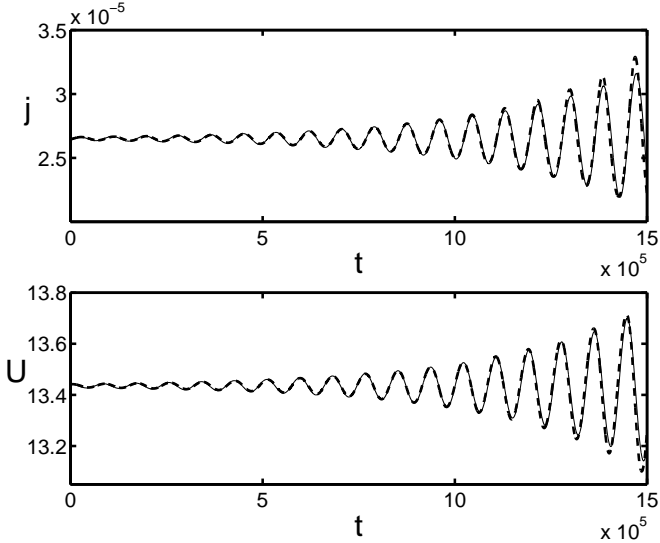}\\
    \caption{Comparison of $j(\tau)$ and ${\cal U}(\tau))$ 
from the stability analysis (solid lines) 
with the result from the simulation (dashed lines) 
for the parameter values of Figs.~5-6 where the stationary
solution is unstable.}
  \end{center}
\end{figure}

Of course, the predictive power of linear stability analysis is limited
to small perturbations with $j_1\ll j_0$ and ${\cal U}_1\ll{\cal U}_0$.
When the amplitude of the oscillation from Fig.~6 increases further,
nonlinear couplings set in and the system finally reaches a limit 
cycle as shown in Fig.~7.

%\vspace{3cm}

%\newpage

\subsection{Calculation of phase diagrams}

The stability analysis now allows one to derive the bifurcation line
where a homogeneous stationary state looses its stability.
Fig.~10 shows this bifurcation line for the parameters (\ref{param})
as a function of applied voltage ${\cal U}_t$ and conductivity $1/{\cal R}_s$
for three different values of $\gamma$. Besides the value 
$\gamma=0.08$ used everywhere else in the paper, also results
for $\gamma=0.04$ and 0.16 are shown to illustrate the sensitivity 
of theoretical predictions to this parameter. 
For Re~$\lambda<0$, the stationary state is linearly stable, 
while for Re~$\lambda>0$, the system is always in the oscillating state.

\begin{figure}[htbp]
  \begin{center}
    \includegraphics[width=0.49\textwidth]{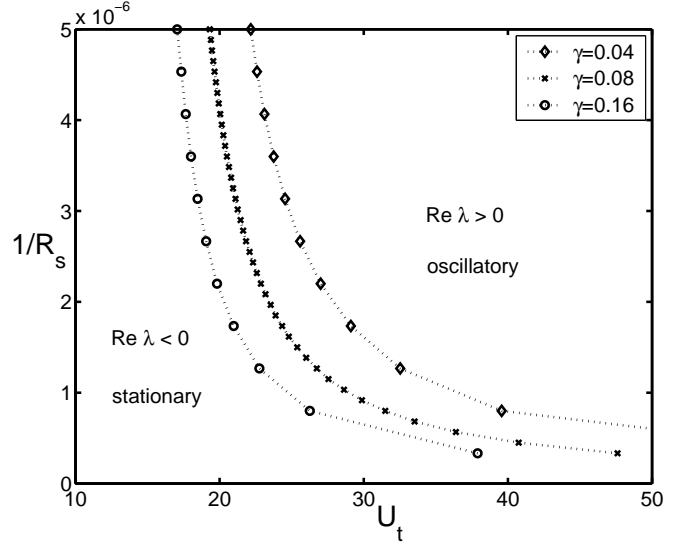}\\
    \caption{Bifurcation diagram for the parameters from (\ref{param}) 
     (where $L=36$) and 3 different values 
        of $\gamma$. The lines separate regions with Re~$\lambda<0$ 
        where the stationary state is linearly stable from 
        regions with Re~$\lambda>0$ where the homogeneous stationary
        state looses its stability. \label{bifdi}}
  \end{center}
\end{figure}

\begin{figure}[htbp]
  \begin{center}
    \includegraphics[width=0.49\textwidth]{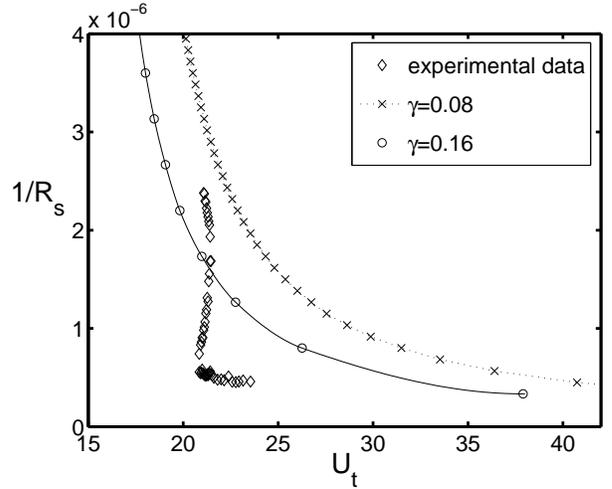} \\
%bifexpM7aR.eps} \\
%expinterp3.eps}\\
    \caption{Blow up of the bifurcation diagram in Fig.~\ref{bifdi} for two different values 
        of $\gamma$ and comparison with experimental data \cite{privatStr,myThesis}. Theoretical lines and experimental lines are in same region of parameters and have same limits.}
  \end{center}
\end{figure}

Comparison with the experimental phase diagram in Fig.~11 for the gas gap with
a corresponding width of $d=1$~mm \cite{privatStr,myThesis} shows qualitative
and quantitative correspondences, but also deviations.
Experiments in the 1 mm gap 
for $U_t<585$~V $({\cal U}_t=20.5)$ do not exhibit oscillations. The same holds theoretically 
 for a secondary emission coefficient of 0.08 or smaller.
In detail, experiments show that the raising 
phase transition line %experimentaly obtained 
initially raises 
with positive slope then changes gradually to being almost parallel to the $\sigma_s$ 
axis and then continues with negative slope up to the maximal experimentally reached  $\sigma_s$.
\iffalse
the left hand side of the diagram is 
straighter than in the 0.5 mm case from \cite{Str}, it raises 
with positive slope from
$(U_t,\sigma_s)=\big(585~{\rm V},~ 0.7\cdot10^{-7}/(\Omega~{\rm cm})\big)$
to $\big(610~{\rm V},~ 2.0\cdot10^{-7}/(\Omega~{\rm cm})\big)$,
where the slope has changed gradually to being parallel to the $\sigma_s$ 
axis. The bifurcation line then continues with negative slope up to 
$\big(600~{\rm V},~ 2.8\cdot10^{-7}/(\Omega~{\rm cm})\big)$.
In dimensionless units, these points on the raising bifurcation line 
are $({\cal U}_t,1/{\cal R}_s)=\big(20.5,~ 0.7\cdot10^{-6}\big)$, 
$\big(21.4,~ 2.1\cdot10^{-6}\big)$ and $\big(21,~ 2.9\cdot10^{-6}\big)$.

The upper part of this experimental transition line is rather well
described by the theoretical curve while %for the somewhat large value $\gamma=0.16$
 in the lower part, the shape of experimental 
and theoretical curve deviate.
\fi

For the low conductivity of the  semiconductor layer, the experiment shows another bifurcation line almost
parallel to the $U_t$ axis at values of $\sigma_s$ %between $0.7\cdot10^{-7}/(\Omega~{\rm cm})$ and 
 around $0.5\cdot10^{-7}/(\Omega~{\rm cm})$.
%that intersects with the raising line. 
In dimensionless units
this corresponds to a plateau at values of $1/{\cal R}_s$ %between $0.7\cdot10^{-6}$ and 
around $0.5\cdot10^{-6}$.
An approach to such a plateau can also be seen in the calculated phase
diagram. However, the theoretical curve crosses over continuously
to this plateau, while the experimental curve seems to show the
intersection of two bifurcation lines with quite distinct slope.
We have no explanation for this deviation.

It is remarkable that the bifurcation theory also covers
the almost horizontal bifurcation line for small $1/{\cal R}_s$.
Another explanation for this experimentally observed feature
of the phase diagram would have been a breakdown of the continuum
approximation: the recovery phase of the oscillation would have
carried such a low current that the discreteness of the electrons
would have to be taken into account.

Finally, it was observed experimentally \cite{Str,privatStr} that increasing 
the system size $L$ while keeping other conditions unchanged, 
the frequency decreases and oscillations set in at higher voltages.
This agrees with our calculated phase diagram in Fig.~11.
Indeed, for ${\cal U}_t<22.5$, the homogeneous stationary state 
is stable for $L=72$. 
%(For L=70, we still see the oscillation (though without separation of time scales), and I'll check L=72) Explanation for that might be that they were looking for them in wrong range of parameters, namely for very small voltages or that all points are just stable)
%Try to get biff.diagram for L=72, same R, and voltages slightly higher then towns...acctually I can always find region where I have only stable solution, so I'll do that and plot that fig as the illustration or not?

\begin{figure}[htbp]
  \begin{center}
    \includegraphics[width=0.49\textwidth]{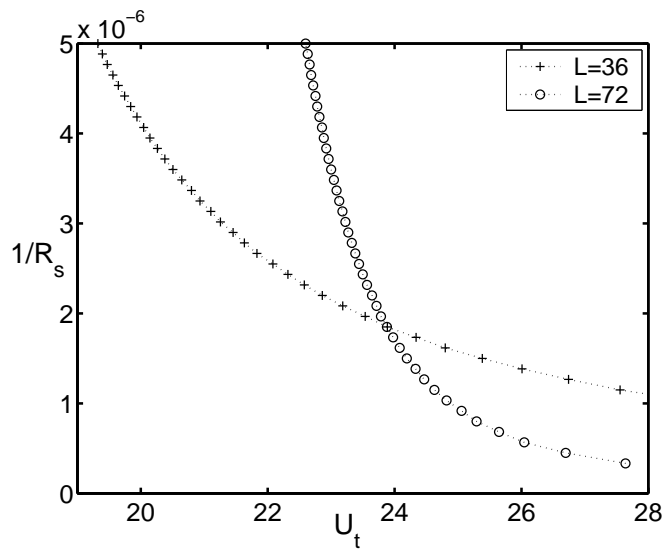}
    \caption{Bifurcation diagram for $\gamma = 0.08$, $L=36$ and $L=72$.}
  \end{center}
\end{figure}

\section{Conclusion}

We have analyzed the simplest model for a one-dimensional short
gas discharge coupled to an external circuit with resistor, 
capacitance and stationary voltage. This analysis is directly 
applicable to experiments performed in \cite{Str,privatStr}.

We have presented fully numerical solutions as well as 
a linear stability analysis of the stationary state of the system
which are in very good mutual agreement. The numerical solutions
reproduce experimental observations of bistability and oscillations
in a semi-quantitative manner, though the model is minimal and no
attempt of parameter fitting has been made.
The stability analysis allows us to derive bifurcation diagrams
in a simple manner, they also agree overall with experimentally
obtained bifurcation diagrams.

It should be remarked that we have constrained the analysis
to the gap of 1 mm wide; the gap of 0.5 mm is so sensitive
to the actual value of secondary emission $\gamma$ that
quantitative analysis based on a fixed value of $\gamma$
seemed doubtful.

We have reproduced a number of experimental
observations up to the dependence of oscillation amplitude 
on applied potential and of the oscillation frequency
on the conductivity of the semiconductor layer,
while discrepancies of other observables will stay a 
subject of investigation. This opens up the way to investigate
now the spatial and spatio-temporal patterns in the next step.

%\newpage

%\vspace{1cm}

{\bf Acknowledgment}: {We acknowledge very useful discussions about 
the experiments with C.\ Str\"umpel, H.-G.\ Purwins, Y.A.\ Astrov
and other members of the M\"unster group. We had useful discussions 
with W.\ Hundsdorfer about numerical solutions and with Yu.P.\ Raizer
about the nature of the oscillations.\\
The work of D.S.\ was supported by the Dutch physics
funding agency FOM, and the work of I.R.\ was made possible
mainly by the European Consortium for Informatics and Mathematics (ERCIM)
and also by FOM.}

\end{document}